\documentclass[12pt,a4paper]{article}

\usepackage{amsmath,amssymb,amsfonts,bm,graphicx,xspace}

\usepackage{enumerate}
\usepackage{cite}
\usepackage{xspace}
\usepackage{ifthen}
\usepackage{url}
\usepackage{booktabs}
\usepackage{multirow}
\usepackage{subcaption}
\usepackage{marginnote}
\usepackage{lineno}
\usepackage{amsmath,bm,graphicx}
\usepackage[utf8]{inputenc}
\usepackage{xcolor}

\newcommand{\pythia}{P\protect\scalebox{0.8}{YTHIA}\xspace}

\newcommand{\pytppp}{P\protect\scalebox{0.8}{YTHIA}8\xspace}

\newcommand{\alice}{ALICE\xspace}

\newcommand{\angantyr}{Angantyr\xspace}

\renewcommand{\eqref}[1]{eq.~(\ref{#1})\xspace}

\newcommand{\fig}[1]{\ref{#1}}
\newcommand{\figref}[1]{figure~\fig{#1}}

\newcommand{\citeref}[1]{ref.~\cite{#1}}
\newcommand{\citerefs}[1]{refs.~\cite{#1}}

\def\text{\mathrm}

\def\eg{\emph{e.g.}\xspace}

\def\mrm#1{\mathrm{#1}}

\def\pp{\ensuremath{\mrm{pp}}\xspace}

\def\pA{\ensuremath{\mrm{p}A}\xspace}

\def\pPb{\ensuremath{\mrm{pPb}}}

\def\AA{\ensuremath{AA}\xspace}

\def\PbPb{\ensuremath{\mrm{PbPb}}}

\def\keff{\ensuremath{\kappa_{\mrm{eff}}}}

\def\newsection#1{~\\ \textit{#1:--}\noindent}

\def\ColText(#1,#2)[#3]#4{\Text(#1,#2)[#3]{#4}}

\def\showcommentsflag{0}
\newcommand{\showcomments}{\def\showcommentsflag{1}}

\newcounter{commentcounter}%
\definecolor{armygreen}{rgb}{0.29, 0.33, 0.13}
\newcommand{\Red}[1]{\textcolor{red}{#1}}
\newcommand{\Black}[1]{\textcolor{black}{#1}}

\newcommand{\comment}[1]{\ifnum\showcommentsflag > 0%
        \addtocounter{commentcounter}{1}%
        {{\Red{\ensuremath{\ddagger^{\arabic{commentcounter}}}}\Black{}}}%
        \marginpar{\raggedright\tiny\it{{\Red{\ensuremath{\ddagger^{\arabic{commentcounter}}}}} {#1}\Black{}}}
        \fi%
}
\newcommand{\commentdel}[2]{\ifnum\showcommentsflag > 0%
        \Red{\sout{#1}}\comment{#2}%
        \fi
}
\newcommand{\commentadd}[2]{\ifnum\showcommentsflag > 0%
        \comment{#2}\Red{#1}%
        \else
        #1
        \fi
}
\newcommand{\commentchange}[3]{\ifnum\showcommentsflag > 0%
        \Red{\sout{#2}}\comment{#3}\Red{#1}%
        \else
        #1
        \fi
}
\newcommand{\nocomment}[1]{\ifnum\showcommentsflag > 0%
        {\tiny\it\Red{\{#1}\}}
        \fi%
}
\newcommand{\nocommentdel}[1]{\ifnum\showcommentsflag > 0%
        \Red{\sout{#1}}%
        \fi
}
\newcommand{\nocommentadd}[1]{\ifnum\showcommentsflag > 0%
        \Red{#1}%
        \else
        #1
        \fi
}

\newcommand{\nocommentchange}[2]{\ifnum\showcommentsflag > 0%
        \Red{\sout{#2}}\Red{#1}%
        \else
        #1
        \fi
}

\showcomments

\newlength{\abstwidth}
\begin{document}
\sloppy
 
\pagestyle{empty}
 
\begin{flushright}
LU-TP-22-25\\
MCnet-22-11\\
\end{flushright}

\vspace{\fill}

\begin{center}
{\Huge\bf Strangeness enhancement across collision systems without a plasma}\\[4mm]
{\Large Christian~Bierlich, Smita~Chakraborty, Gösta~Gustafson, and Leif~Lönnblad} \\[3mm]
{\texttt{christian.bierlich@thep.lu.se}, \texttt{smita.chakraborty@thep.lu.se}, \texttt{gosta.gustafson@thep.lu.se}, \texttt{leif.lonnblad@thep.lu.se}}\\[1mm]
{\it Theoretical Particle Physics,}\\[1mm]
{\it Department of Astronomy and Theoretical Physics,}\\[1mm]
{\it Lund University,}\\[1mm]
{\it S\"olvegatan 14A,}\\[1mm]
{\it SE-223 62 Lund, Sweden}
\end{center}

\vspace{\fill}

\begin{center}
  \begin{minipage}{\abstwidth}
    {\bf Abstract}\\[2mm]
    We present novel \textit{rope} hadronization results for strange
    hadron enhancement in \pp, \pPb, and \PbPb\ collisions using
    \pythia/\angantyr at LHC energies. With the rope model for string
    fragmentation, we find that the strangeness and baryon enhancement
    has a coherent increase across all collision systems as a function
    of average charged central multiplicity, in qualitative agreement
    with LHC data. In \AA\ collisions, we find that the baryonic
    yields overshoot data at high multiplicities, and we discuss how
    a combination of rope hadronization with other string interactions
    may tame this rise.
\end{minipage}
\end{center}

\vspace{\fill}

\phantom{dummy}

\clearpage

\pagestyle{plain}
\setcounter{page}{1}

Strangeness enhancement in both small and large systems is usually
interpreted as a signal of a dense and hot Quark-Gluon Plasma (QGP)
\cite{Rafelski:1982pu, ALICE:2013xmt, ALICE:2016fzo}. However, the Lund strings physically represent colour-electric flux tubes, and
in a series of papers
\cite{Bierlich:2014xba,Bierlich:2017vhg,Bierlich:2020naj,Bierlich:2022oja}
we have demonstrated that interactions between overlapping flux
tubes are able to reproduce not only enhanced rates for
strangeness and baryons \cite{Bierlich:2014xba}, but also collective
flow \cite{Bierlich:2020naj} in \pp\ collisions. The Angantyr model \cite{Bierlich:2018xfw}, which is a generalization
of the Lund string model in \pythia to nucleus-nucleus (\AA)
collisions, is able to successfully reproduce many features of hadron
production in these collisions. We note that in this picture the initial energy density and
temperature do not have to be very high \cite{Bierlich:2020naj}, and
that the string degrees of freedom therefore survive the initial,
mainly longitudinal expansion, until hadronization sets in. In a dense environment, the overlap among these transversely extended
strings naturally causes interactions between overlapping flux
tubes. 

The string-string interaction can be a repulsive interaction between a
pair of overlapping flux tubes, called string shoving
\cite{Bierlich:2017vhg,Bierlich:2020naj}, giving rise to a
transverse collective flow in high-density systems.  The interaction
can also give the formation of ``colour ropes'' as discussed in
\citeref{Biro:1984cf}. The increased energy density in a rope implies
that more energy is released when a new $q\bar{q}$ pair is produced
(or a diquark-antidiquark pair in case of baryon production). This
corresponds to a higher effective string tension $\kappa$ (\keff)
during the hadronization of a rope, which modifies the fragmentation
parameters entering the Lund fragmentation function
\cite{Bierlich:2014xba, Bierlich:2022vdf, Andersson:1983ia}. Quark-antiquark production in a colour-electric field can be regarded
as a tunnelling process \cite{Brezin:1970xf}, which gives a production
probability for different quark flavours 
proportional to $\exp (-\pi \mu^2/ \kappa)$, where $\mu$ is the respective (di-)quark mass.  An increased $\kappa$ will here mainly
reduce the suppression of strange quarks (and diquarks), while heavier
quarks typically are too suppressed to be relevant for the
hadronization process, and are only produced in hard scattering
processes. Thus the higher \keff\ produces higher yields of both
strange hadrons and baryons in \pp\ collisions, as described in detail in
\citeref{Bierlich:2014xba}.
To use rope hadronization in \angantyr-generated \pA\ and \AA\
collisions, we require a common reference frame for every possible
pair of strings. Such a reference frame would be a baseline for
computing the interactions between every possible string
pair. In this paper, we use a new implementation of the rope model, based
on this so-called \emph{parallel} frame \cite{Bierlich:2022oja}, and
show the resulting strangeness enhancement in \pp, \pPb, and \PbPb\
systems. We conclude by examining our current approach and outlining
possible improvements.

\newsection{Strangeness enhancement due to rope hadronization} Our
original rope hadroni\-zation model, and the recent re-implementation
of it, are presented in detail in \citerefs{Bierlich:2014xba} and
\cite{Bierlich:2022oja} respectively. When we now apply the model to
collisions involving heavy ions we use the \angantyr
\cite{Bierlich:2018xfw} model. Parton-level nucleon--nucleon
collisions generated by \pythia are here stacked on top of each other,
after having selected nucleon sub-collisions using a
Glauber simulation. When hadronizing a given set of such
sub-collisions, it is in principle straight forward to apply our rope
model, but there are a few caveats, and the main one of these is
related to \emph{colour reconnections} (CR). To understand this issue
we need go through some of the basics of the rope model.

Consider two simple strings, each stretched between a quark and an
antiquark. If they are completely overlapping and anti-parallel, the
colours of the quark and anti-quark in each end can either combine
into a colour-octet or a colour-singlet. Lattice calculations show
\cite{Bali:2000un} that the tension in a \emph{rope} between any
multiplet and the corresponding anti-multiplet charge is proportional
the second Casimir operator. Denoting a multiplet corresponding to $p$
coherent triplet and $q$ coherent anti-triplet colours by $\{p,q\}$,
this gives
\begin{equation}
  \label{eq:ropetension}
  \kappa^{\{p,q\}}/\kappa^{\{1,0\}}=C_2^{\{p,q\}}/C_2^{\{1,0\}}=
  \frac{1}{4} \left(p^2 + pq + q^2 + 3p + 3q\right),
\end{equation}
where $\kappa^{\{1,0\}}\equiv\kappa$ is the tension in a single
triplet string. Thus for an octet, we would have a
string tension of $9\kappa/4$, while for a singlet, we would have no string at
all. In pp collisions, all strings originate in a small region in coordinate space,
and we have previously argued in \citeref{Bierlich:2022oja} that
singlet case corresponds to the CR process in \pythia
\cite{Sjostrand:1987su}, where strings close in momentum space
are allowed to reconnect, if the combined string length is thereby
reduced. For high string densities we assumed that any combination
of $m$ triplets and $n$ anti-triplets would then always combine into the
highest possible multiplet \cite{Bierlich:2022oja}.

For nuclear collisions, we must also account for the separation between
vertices in coordinate space.  In the \angantyr model, although there
are CRs between strings formed in the same sub-collision, there are
currently no reconnections between strings from different
sub-collisions, since the current CR models in \pythia do not take
into account the space-time separation.

To handle the space-time separation for rope formation, we here use
the \emph{parallel frame formalism} \cite{Bierlich:2022oja} to include
separation between vertices in both coordinate and momentum space. It
also accounts for strings with ``kinks'', coming from strings
stretched between a quark-antiquak pair via a number of gluons.  It
constructs a Lorentz transformation to a frame, where any two
(straight) string pieces will always lie in parallel planes, moving
away from each other with equal velocities.

To estimate the \emph{formation of a rope} in the absence of CR, we now
adopt a \emph{random walk} in colour space combining elementary colour
charges \cite{Biro:1984cf} where we add one overlapping string at the time. Adding a triplet to a
$\{p,q\}$ multiplet can give three possibilities:

\begin{eqnarray}
  \{p,q\} \oplus \{1,0\} & = &\left\{
    \begin{array}[c]{lr}
      \{p+1,q\} &  \\ 
      \{p,q-1\} &  \\ 
      \{p-1,q+1\} & 
    \end{array}\right.
	\label{eq:randomwalk}
\end{eqnarray}
and similarly for adding an anti-triplet. For each added overlapping string, we choose randomly among these, according to the number of
states in the corresponding multiplet,  $N_{\{p,q\}} = (p+1) (q+1) (p+q+2)/2$. Since we always assume that we have a string being hadronized ($\{1,0\}$)
to begin with, we never let $p$ go to zero, but otherwise the random
walk is unconstrained.

To estimate the \emph{breakup of a rope} we note that the tension 
in a rope $\{p,q\}$ is given by the Casimir operator in \eqref{eq:ropetension}. In our model, the rope would break up one string at the
time, and the tunnelling process implies that in each such a breakup, the 
effective string tension is given
by the \emph{reduction} of the rope tension due to the multiplet
field being reduced from, \eg, $\{p,q\}\mapsto\{p-1,q\}$:
\begin{equation}
  \label{eq:kappaeff}
  \keff=\kappa^{\{p,q\}}-\kappa^{\{p-1,q\}}=\frac{2p+q+2}{4} \kappa.
\end{equation}

In our new implementation, for any given breakup in a string being
hadronized, we can consider each of the other string pieces in the
event, and boost to the common parallel frame for this pair. Assuming
that the dominant part of the string interaction is given by the
colour-electric field, using an ``Abelian projection'' of the
SU(3) field [9, 10], the total interaction is the sum of the
interaction between all pairs
($\int d^3x\,\mathbf{E}_{\mathrm{tot}}^2 = \sum_{i,j} \int d^3x\,
\mathbf{E}_i \mathbf{E}_j$). In the parallel frame, we can estimate
the fractional overlap based on their space-time location, their
relative angle, and assuming a Gaussian transverse shape as described
in detail in \citeref{Bierlich:2022oja}. These fractional overlaps can
then be added together to give the numbers of steps $m$ and $n$ in the
random walk to obtain the multiplet $\{p,q\}$. (In contrast to the
highest-multiplet procedure used in \cite{Bierlich:2022oja} and in
\figref{fig:lambda-k-ratio} below, where we set $p=m,\,q=n$.)

In each of these breakups we can determine an effective string
tension, \keff. This will influence the flavour of the $q\bar{q}$ pair
responsible for the breakup via the tunnelling mechanism, notably
increasing strange quark production. Also di-quarks can be produced in
the breakups, giving rise to baryon production, and especially strange
baryons are enhanced by an increase in \keff. These effects are
technically achieved by changing several parameters in \pytppp, as
explained in \citeref{Bierlich:2014xba}.

\begin{figure}
	\centering
  \includegraphics[scale=0.9]{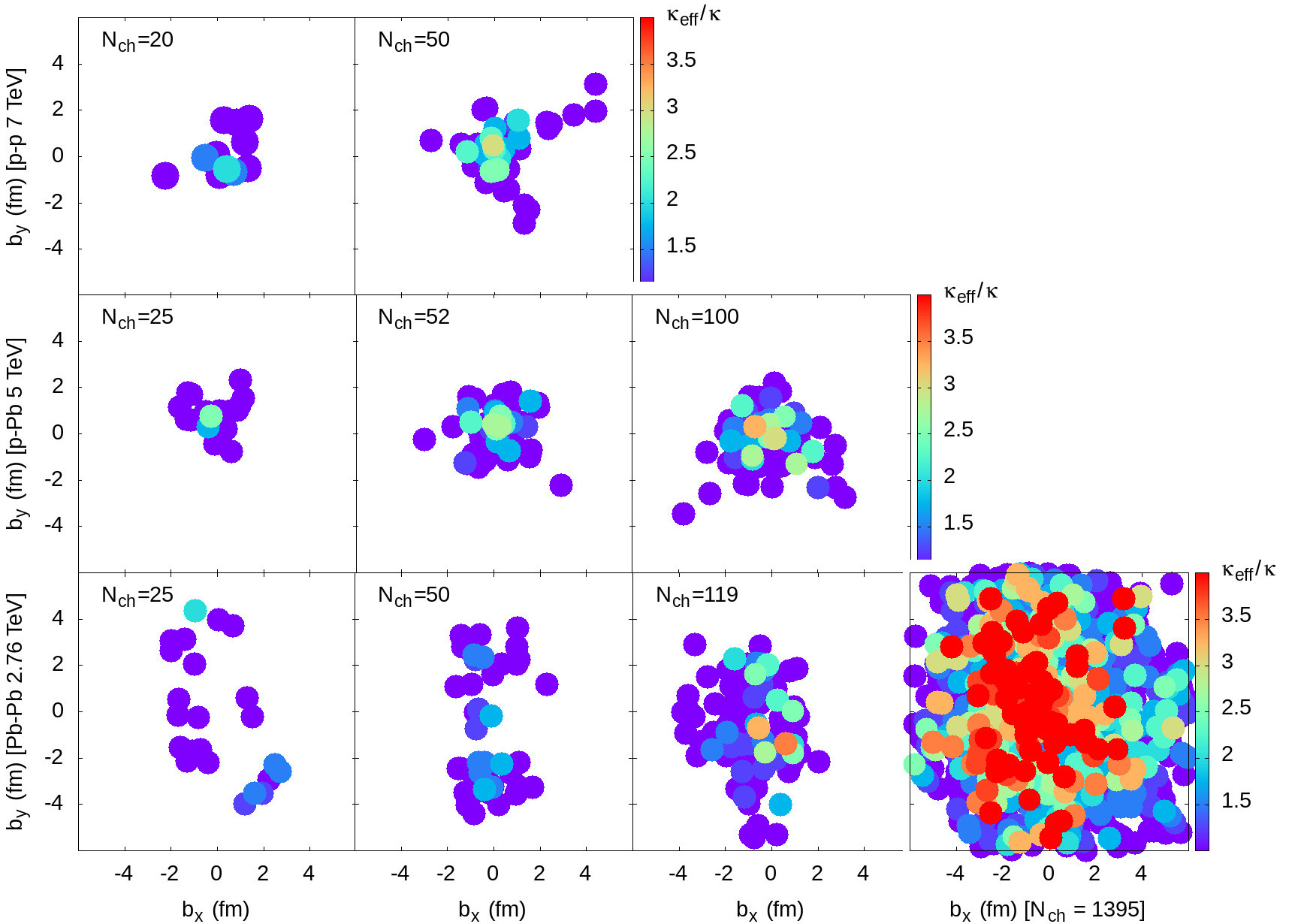}
  
  \caption{Production points of primary hadrons in impact parameter
    space produced in the central pseudo-rapidity bin in sample events
    from \pp\ at 7~TeV (top row), \pPb\ at 5.02~TeV
    (middle), and \PbPb\ collisions at 2.76~TeV (bottom) for different
    intervals of central ($|\eta| < 0.5$) charged multiplicity:
    $\sim25$ (first column), $\sim50$ (second), $\sim100$ (third) and
    $>$1000 (last column). The colour of the points indicates the
    $\keff/\kappa$ used in the string breakup where the primary
    hadrons were produced. The impact parameter vector is along the
    $x$-axis.}
 \label{fig:sys-scat}
        
\end{figure}
\def\dndeta{\ensuremath{\left.dN_{\text{ch}}/d\eta\right|_{\eta=0}}}
\def\avdndeta{\ensuremath{\left\langle dN_{\text{ch}}/d\eta\right\rangle_{\eta=0}}}

To illustrate the variations in \keff, 
we show in \figref{fig:sys-scat} sample events from 
\pp, \pPb, and \PbPb\ collisions.  Each point in each figure 
represents the production point in
impact parameter space of a primary hadron (a hadron produced
directly in the hadronization) produced in the central unit of
pseudo-rapidity, and the colour indicates the $\keff/\kappa$ used to produce
it. The size of the points correspond to the assumed width of the
strings ($R\approx0.5$~fm). Since the number of produced hadrons
per unit rapidity is approximately the same in all vertical columns,
the figure also gives an indication of the difference in (transverse)
density of strings for different systems.

We show events with a central charged
multiplicity%
\footnote{Note that $N_{\text{ch}}$ is not the same as the number of primary hadrons.},
$\dndeta\sim25$ (leftmost column), for \pp,
\pPb, and \PbPb, and we see that the \PbPb\ event is much
more spread out, as expected. We note that also the \pp\ event is
quite spread out. This is because (mini-)jets cause the hadron
production to occur away from the centre of the collision. This is in
line with the expectation that the increase in multiplicity in \pp\
collisions is driven by additional jet production while in \pA,
and especially in \AA, the increase is due to additional
nucleon--nucleon sub-collisions. Although there is only one event from
each collision system, we see that there is no dramatic difference in
the average \keff. Also for $N_{\text{ch}}\sim50$ (second
column), we see no large difference in \keff, and while we see the
expected almond-like shape for \PbPb, the \pp\ event is
clearly more \emph{jetty}. At $N_{\text{ch}}\sim110$ (third column), there are fewer \pp\ events, and we only show samples from
\pPb\ and \PbPb. Here we see that the \pPb\ has jets, but the
difference between \keff\ in \pPb\ and \PbPb\ is still not
large. To get higher \keff, we need extremely high
multiplicities, available only in \PbPb\ collisions, and we show such
an event in the last column. The density of strings is here
very large, but there are also large fluctuations in
density, with \textit{hotspots} spread out in the impact parameter
plane.

\begin{figure*}
	\centering
	\includegraphics[scale=0.7]{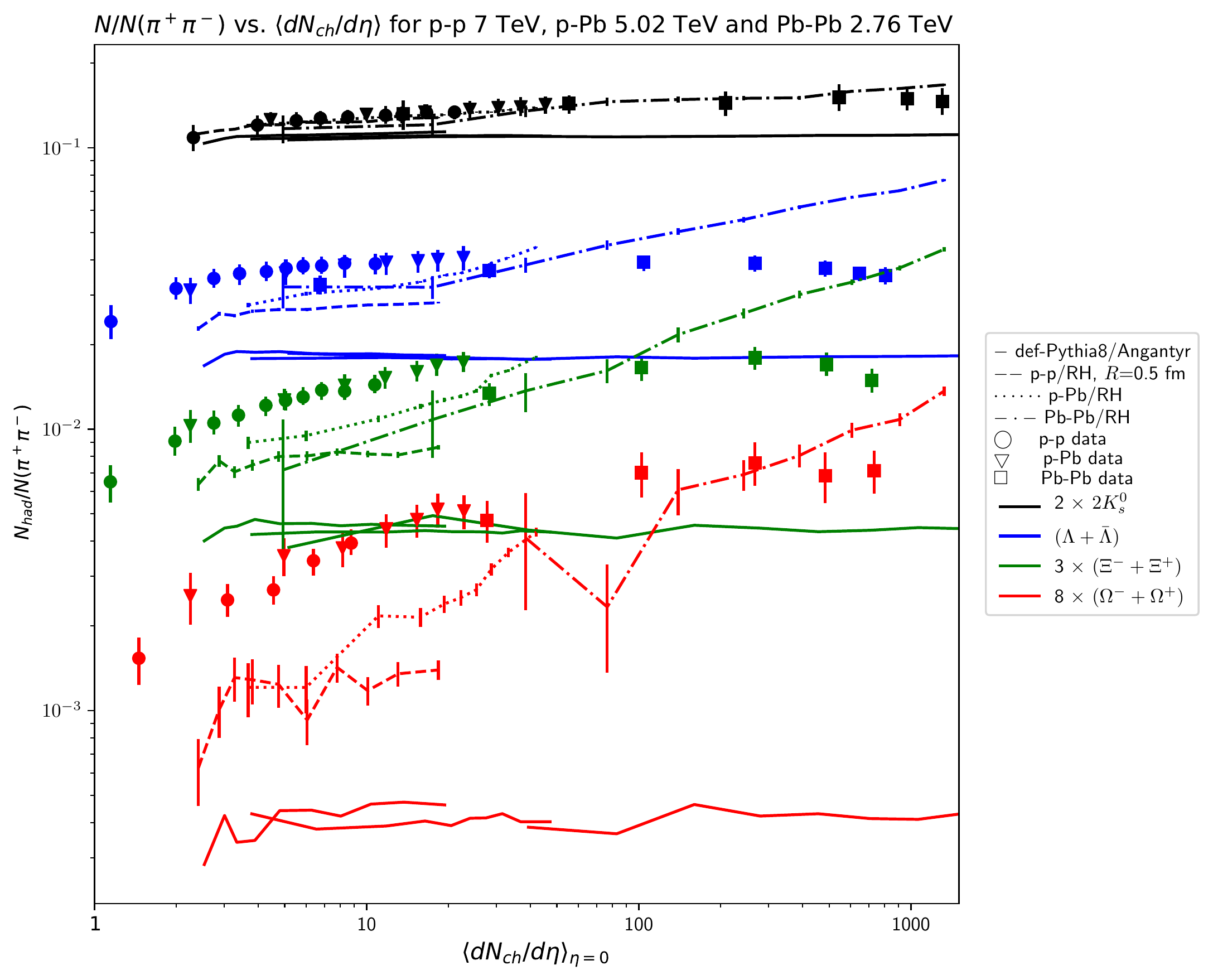}
	\caption{Strange hadron to pion yield ratios in \pp\ at
          $\sqrt{s} = 7$~TeV, \pPb\ at
          $\sqrt{s_{\text{NN}}} = 5.02$~TeV, and \PbPb\ at
          $\sqrt{s_{\text{NN}}} = 2.76$~TeV, vs.\
          \avdndeta. The data is taken from \alice
          \cite{ALICE:2016fzo}. For clarity error bars are only shown for the rope model results.}
	\label{fig:enhance}
\end{figure*}

\newsection{Results}
We now apply the rope hadronization mechanism to minimum bias events
in the three systems with a canonical value of the string width $R=0.5$~fm. The analysis for each system follows the procedure used by \alice in
\citeref{ALICE:2016fzo}, where the event samples are divided in
centrality bins, and the the ratio of each of the strange hadrons to
the pion rate are presented as a function of the average central
charged multiplicity, \avdndeta, in
each bin. We have used default \pythia/\angantyr, and the tuned
hadronization parameters presented in \citeref{Bierlich:2022vdf}.
In \figref{fig:enhance} we show our result in all three systems,
with and without rope hadronization compared to data
\cite{ALICE:2016fzo}.

We look at $K^0_s$, $\Lambda$, $\Xi$ and $\Omega$ hadrons to $\pi$
ratio for $|\eta| < 0.5$. Overall, the yields of strange mesons and
baryons are significantly increased due to rope formation compared to the baseline prediction of Lund string hadronization without any rope effects, which  is a major improvement. The magnitude of enhancement is directly related to
string density before hadronization sets in, giving an increased
\keff. The number of strings, in turn, is directly related to the average
charge multiplicity at mid-rapidity,\avdndeta. Therefore, \dndeta\ is a perfect scaling variable
to show the multiplicity dependence of \keff, and hence also for strangeness
and baryon enhancement. This scaling is in qualitative agreement with
the data, and for $K^0_s$ we also have quantitative
agreement. However, we see that the increase with \avdndeta\ is
too steep, especially for the baryons, but also for $K^0_s$ there is a
tendency to overshoot the data at the highest multiplicity in
\PbPb. In our results for the rope model we have assumed $R=0.5$~fm
for the string radius, and increasing it to 1~fm would improve the
comparison with data for low multiplicities (see
\figref{fig:lambda-k-ratio}), but would overshoot even more at the highest
multiplicity \PbPb\ bins.

We notice that the yield ratios in \pPb\ and \PbPb\ lack saturation
at high \dndeta, while \pp\ is much flatter. As discussed above, this
can be expected since very high multiplicities in \pp\ are, to a large
extent, driven by increased (mini-)jets production, while the high
multiplicities in \AA\ are driven by an increase in soft nucleon
sub-collisions. Since hadrons that are produced in jets come from
strings that are not parallel to the bulk of the soft strings along
the beam axis, and are produced further away from these, the effective
number of overlaps is smaller.

From \figref{fig:enhance}, it may seem that the steep
rise in yield ratios is mainly a problem for the baryon production, but this is not
necessarily the case. In \figref{fig:lambda-k-ratio}, where we show the
$(\Lambda +\bar{\Lambda})/2K^0_s$ ratio for \pp\ at $\sqrt{s}=7$~TeV,
we observe that the increase in $\Lambda/K$ follows the behaviour seen
in data.  There are, however, additional uncertainties in the string
fragmentation for baryons. One such effect is the hyperfine correction
\cite{Bierlich:2022vdf} arising from spin-spin interactions between a
quark and an antiquark or between two quarks. This especially affects the
multi-strange particles, and there is room for further corrections to
the baryon yields. However, this would mainly affect the overall yield of (multi) strange
baryons, and would not affect the strong rise in \figref{fig:enhance},
which is mainly driven by \keff.

Including colour reconnections between sub collisions, which are
lacking in our current implementation will clearly affect the
results. This would naively reduce the number of strings ($n_s$), but
it would not necessarily reduce the rise of \keff\ since
$N_{\text{ch}}$ is approximately proportional to $n_s$. Hence, $n_s$
would have to be increased again by modifying \angantyr parameters in
order to fit data. Additionally, after that we would need to use the
highest multiplet procedure which would increase \keff, although we
see in \figref{fig:lambda-k-ratio} that this is not necessarily a
large effect.

There is, however, one mechanism that can decrease the string density
without decreasing $n_s$, and that is the repulsion
between overlapping strings, which is addressed in the string shoving
model \cite{Bierlich:2017vhg,Bierlich:2020naj}.  Owing to the
technical difficulties outlined in \citeref{Bierlich:2022oja} string
shoving is not included in our current results. However,
shoving would spread out the strings more in dense environments,
possibly taming the rise for high multiplicities in
\figref{fig:enhance}.

\begin{figure}
	\centering
	\includegraphics[scale=0.7]{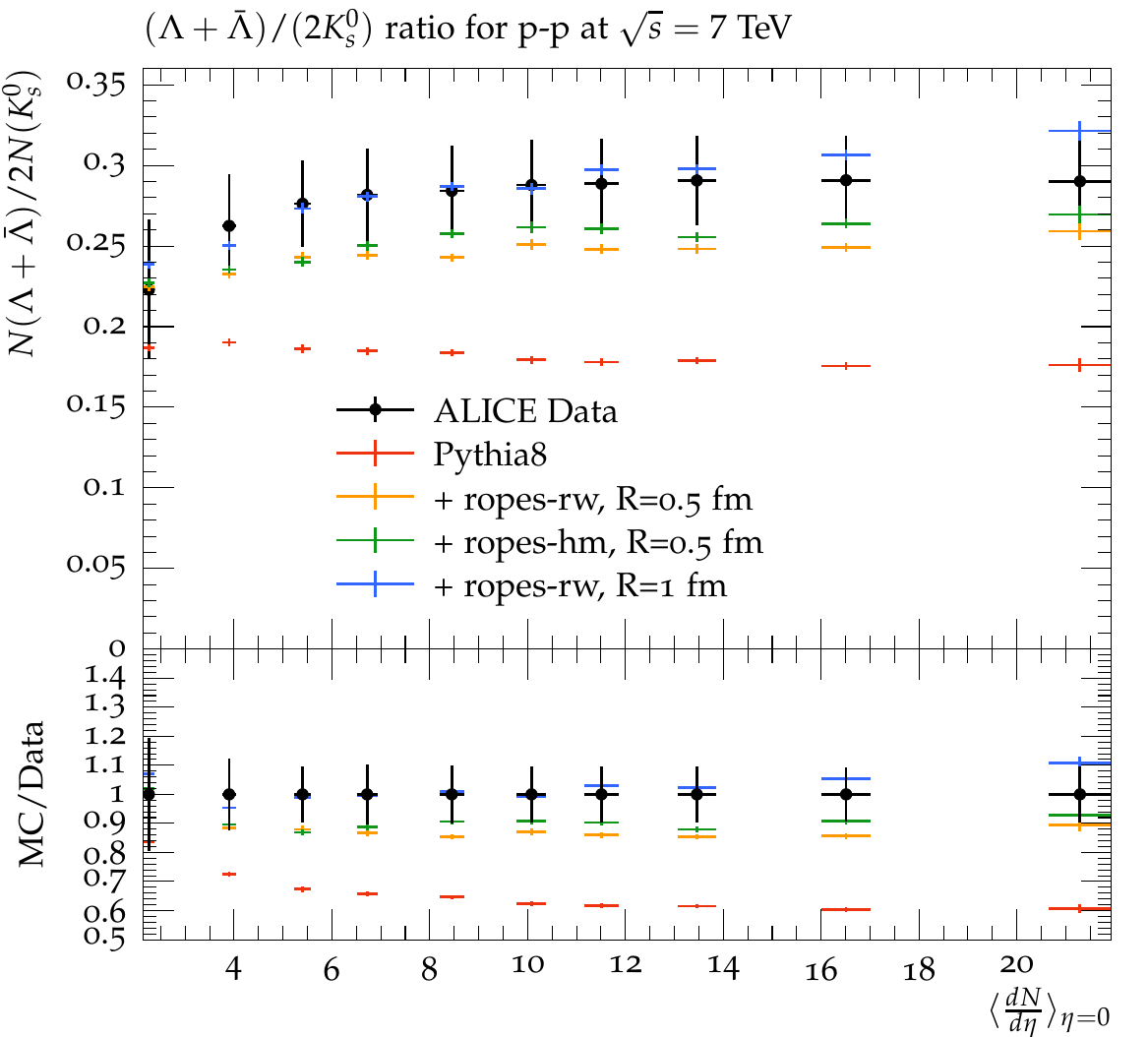}
	\caption{$(\Lambda +\bar{\Lambda})/2K^0_s$ yield ratio vs.\
          \avdndeta, compared to \pp\ minimum bias data at
          $\sqrt{s}=7$~TeV \cite{ALICE:2016fzo}. Effects of random
          walk with $R=0.5$~fm and $1$~fm, and highest-multiplet with
          $R=0.5$~fm is shown.}
	\label{fig:lambda-k-ratio}
\end{figure} 

To look further into rope effects on the baryon yield, we show in
\figref{fig:lambda-k-ratio} the rope model for two values of the string radius with the
random walk (rw) formulation. We also show the yield ratio using
highest multiplet (hm) formulation for $R=0.5$~fm, which we use in
\citeref{Bierlich:2022oja}.  As discussed above, the highest multiplet
is used together with CR in \pp, assuming that the latter corresponds to the
steps downward in \eqref{eq:randomwalk}.
 
In \figref{fig:lambda-k-ratio}, compared to default \pytppp, rope
hadronization with random walk and $R=0.5$~fm enhances the baryon vs.\
meson yields significantly in all multiplicity bins. However,
following the highest multiplet procedure, the yield ratios are
slightly higher, giving a better agreement with data.
Here we note
that with the random walk formulation, if $R$ has a higher value such
as $1$~fm, the effect is in better agreement to data as shown in the
figure. Therefore, we conclude that string shoving and CR would have non-trivial effects
on strangeness and baryon yields, especially at high the highest
string densities.

\newsection{Conclusions}
Based on the success of our previous rope model in explaining
strangeness enhancement in small systems \cite{ALICE:2016fzo}, we show
here that our new implementation, based on the parallel frame, can
model enhanced strange flavour production in \emph{all} collision
systems. The enhancement of strange hadrons shown here is due to
modification of the string tension \keff\ in dense environments. The
average \keff\ in string fragmentation, as shown in
\figref{fig:sys-scat}, combines the effects of local fluctuations in
overlap among ropes during hadronization. Clearly the current rope model lacks the saturation at high
multiplicities seen in data, but we believe that this is due to the lack of repulsion between
overlapping strings, which we can achieve by combining the ropes with
our shoving model. It should be noted that our rope model is very different from the
conventional picture based on the formation of a QGP, so even if the
same strangeness enhancement can be achieved in both pictures, there
are several other observables that would differ. In particular, it is
central to the string model that there is a strong momentum
correlation between strange and anti-strange hadrons, which should be
completely lacking in a thermalized QGP.

String shoving would not only dilute a string system before
hadronization via rope formation takes place, it can also give rise
to final-state collective flow. Hence, in small systems, both string shoving and rope hadronization
together give rise to two out of three typical QGP-like signals. On the other hand, hadronic rescattering \cite{Sjostrand:2020gyg,
  Bierlich:2021poz} also gives rise to final-state collectivity for
central collisions in \pA\ and \AA. In addition, modification of jet energy
and topology can arise due to CR. Therefore, to arrive to a complete
string-based physical picture in \AA\ collisions in \pythia/\angantyr,
the combined effects of colour reconnections, string shoving, rope
hadronization and hadron rescattering need to be included. Such a
picture could then also be applied to other kinds of collision experiments,
such as in cosmic ray showers, and not least at the future
Electron--Ion collider.

\newsection{Acknowledgements}
This work was funded in part by the Knut and Alice Wallenberg
foundation, contract number 2017.0036, Swedish Research Council, contracts
number 2016-03291, 2016-05996 and 2017-0034, in part by the European
Research Council (ERC) under the European Union’s Horizon 2020
research and innovation programme, grant agreement No 668679, and in
part by the MCnetITN3 H2020 Marie Curie Initial Training Network,
contract 722104.

\bibliographystyle{utphys}
\bibliography{bibliography}

\end{document}